\documentstyle[aps,prb,,epsf,graphicx,amsmath]{revtex}

\begin{document}
\draft
\voffset=2cm

\title{Granular flow down a rough inclined plane: transition between
  thin and thick piles}

\author{Leonardo E.~Silbert, James W. Landry, and Gary S.~Grest}

\address{Sandia National Laboratories, Albuquerque, New Mexico 87185}


\maketitle
\begin{abstract}
  The rheology of granular particles in an inclined plane geometry is
  studied using molecular dynamics simulations. The flow--no-flow
  boundary is determined for piles of varying heights over a range of
  inclination angles $\theta$. Three angles determine the phase
  diagram: $\theta_{r}$, the angle of repose, is the angle at which a
  flowing system comes to rest; $\theta_{m}$, the maximum angle of
  stability, is the inclination required to induce flow in a static
  system; and $\theta_{max}$ is the maximum angle for which stable,
  steady state flow is observed. In the stable flow region
  $\theta_{r}<\theta<\theta_{max}$, three flow regimes can be
  distinguished that depend on how close $\theta$ is to $\theta_{r}$:
  i) $\theta>>\theta_{r}$: Bagnold rheology, characterized by a mean
  particle velocity $v_{x}$ in the direction of flow that scales as
  $v_{x}\propto h^{3/2}$, for a pile of height $h$, ii)
  $\theta\gtrsim\theta_{r}$: the slow flow regime, characterized by a
  linear velocity profile with depth, and iii)
  $\theta\approx\theta_{r}$: avalanche flow characterized by a slow
  underlying creep motion combined with occasional free surface events
  and large energy fluctuations. We also probe the physics of the
  initiation and cessation of flow. The results are compared to
  several recent experimental studies on chute flows and suggest that
  differences between measured velocity profiles in these experiments
  may simply be a consequence of how far the system is from jamming.

\end{abstract}

\section{\label{sec:introduction}Introduction}
The flow properties of granular materials puzzle engineers and
physicists alike. Despite numerous experimental studies of heap flows,
avalanches, and chute flows, there still does not exist a general
consensus on the nature of {\it typical} granular flow. Early
systematic experiments in inclined plane geometries were concerned
with flow rates of granular material down chutes
\cite{savage1,campbell1,campbell2}, usually with smooth bases. Large
regions of plug flow were observed, i.e. the whole bulk of the
material slid down the plane at a velocity determined by the input
mass rate and inclination angle of the chute. Kinetic theories for the
inclined plane problem can explain some aspects of this flow behaviour
for nearly elastic particles on smooth chutes
\cite{brennen1,jackson1}. Other studies have determined structural
features in the flow in two dimensional (2D) experiments \cite{drake1}
and simulations \cite{poschel1,zheng1}.  Still, there has not been the
same success in predicting flows down rough inclines for inelastic,
frictional particles.

Detailed measuring techniques have been developed recently to capture
various flow properties of granular material down rough inclined
planes \cite{pouliquen1,azanza1,ancey2} and in heap flow geometries
\cite{durian1,komatsu1,khakhar2}. Accompanying these more thorough
experiments, further theoretical treatments have also emerged
\cite{azanza1,ancey1,mills4,gollub2,aranson1}, together with full
three dimensional (3D) computer simulations
\cite{walton2,walton3,deniz2,leo7}. Despite this growth in literature,
there appears to be a disturbing feature of these studies: reports on
quantities such as velocity profiles differ quite dramatically. While
some reports are agreement with the pioneering results of Bagnold
\cite{bagnold1,bagnold3}, others are not.

In 1954, Bagnold \cite{bagnold1} proposed a momentum transfer picture
to describe his experimental results on inertial granular flow. This
simple 2D model describes the collisional process between particle
layers during flow. The resulting relations predict a shear stress
$\sigma$ that is proportional to the square of the strain-rate,
\begin{equation}
\sigma \propto \dot{\gamma}^{2},
\label{chute3_eq1}
\end{equation}
where the strain-rate is defined as the derivative of the velocity in
the direction of flow ($x$) with respect to depth ($z$),
$\dot{\gamma}\equiv \partial v_{x}(z)/\partial z$. Typically, in
local, Newtonian, hydrodynamic treatments of granular dynamics
\cite{gollub2}, the relationship $\sigma=\eta\dot{\gamma}$ is
used, combined with the kinetic term $\eta\propto T^{1/2}$, where
$\eta$ is a viscosity and $T$ is the granular temperature (to be
discussed in section \ref{depth}). Since the strain-rate is also
proportional to the square root of the granular temperature
$\dot{\gamma}\propto T^{1/2}$, Eq.~\ref{chute3_eq1} is then recovered.
Bagnold's argument, applied to the case of bulk granular flow, is
predicated on a constant density depth-profile with no-slip condition
at the base. This leads to a velocity depth-profile of the form,
\begin{equation}
v_{x}(z)\propto \left[h^{3/2}-\left(h-z\right)^{3/2}\right],
\label{chute3_eq2}
\end{equation}
for a pile of total height $h$, where $z$ is measured from the bottom
of the pile. Because Bagnold scaling comes naturally from dimensional
analysis in the absence of intrinsic time scales, one might expect that
most studies would observe this kind of rheology. However, results
depend strongly on the experimental procedure; profiles of this form
have only been reported in a few instances \cite{pouliquen1}.

\begin{figure}
\begin{center}
  \includegraphics[width=6cm]{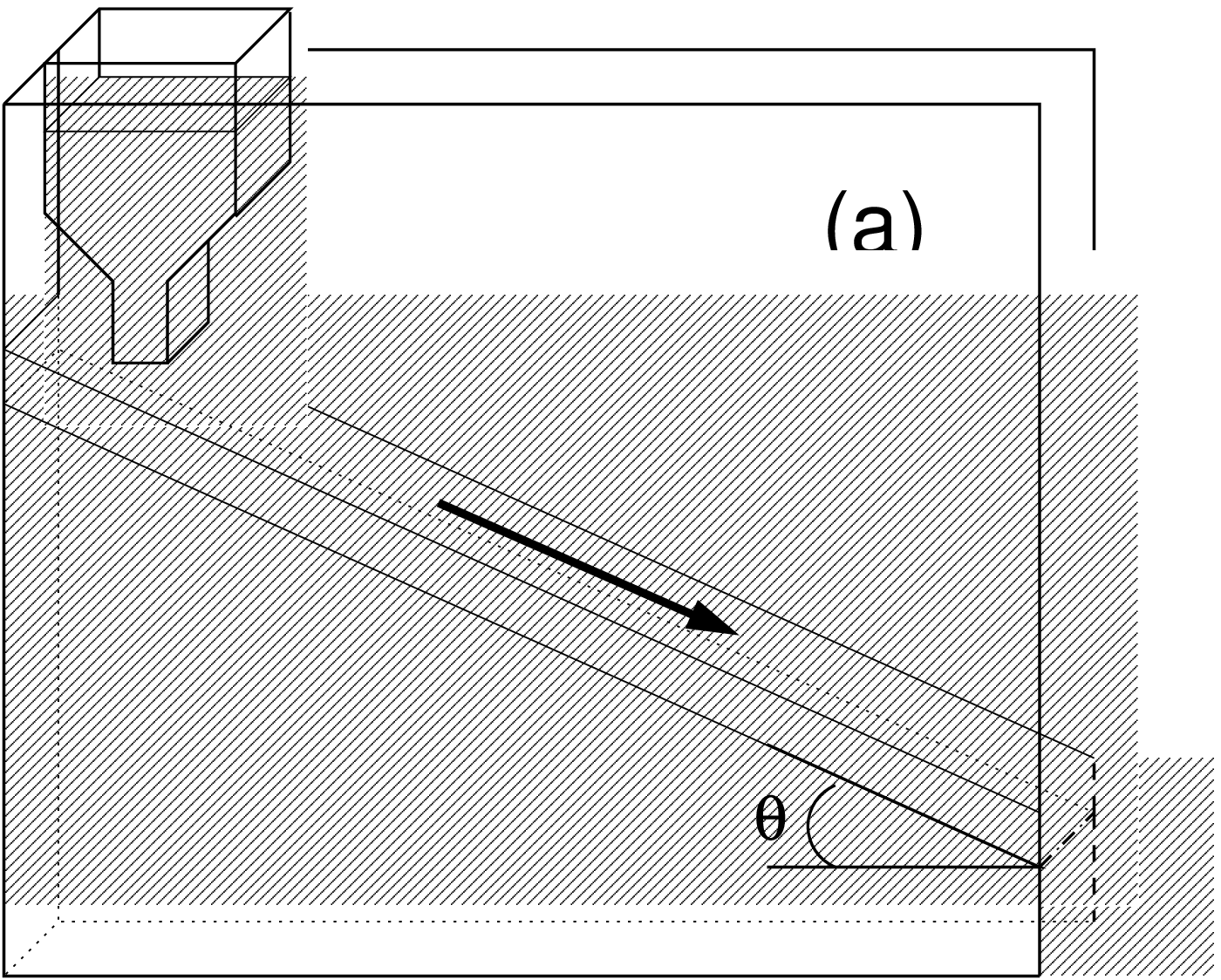}\hspace{2cm}
  \includegraphics[width=6cm]{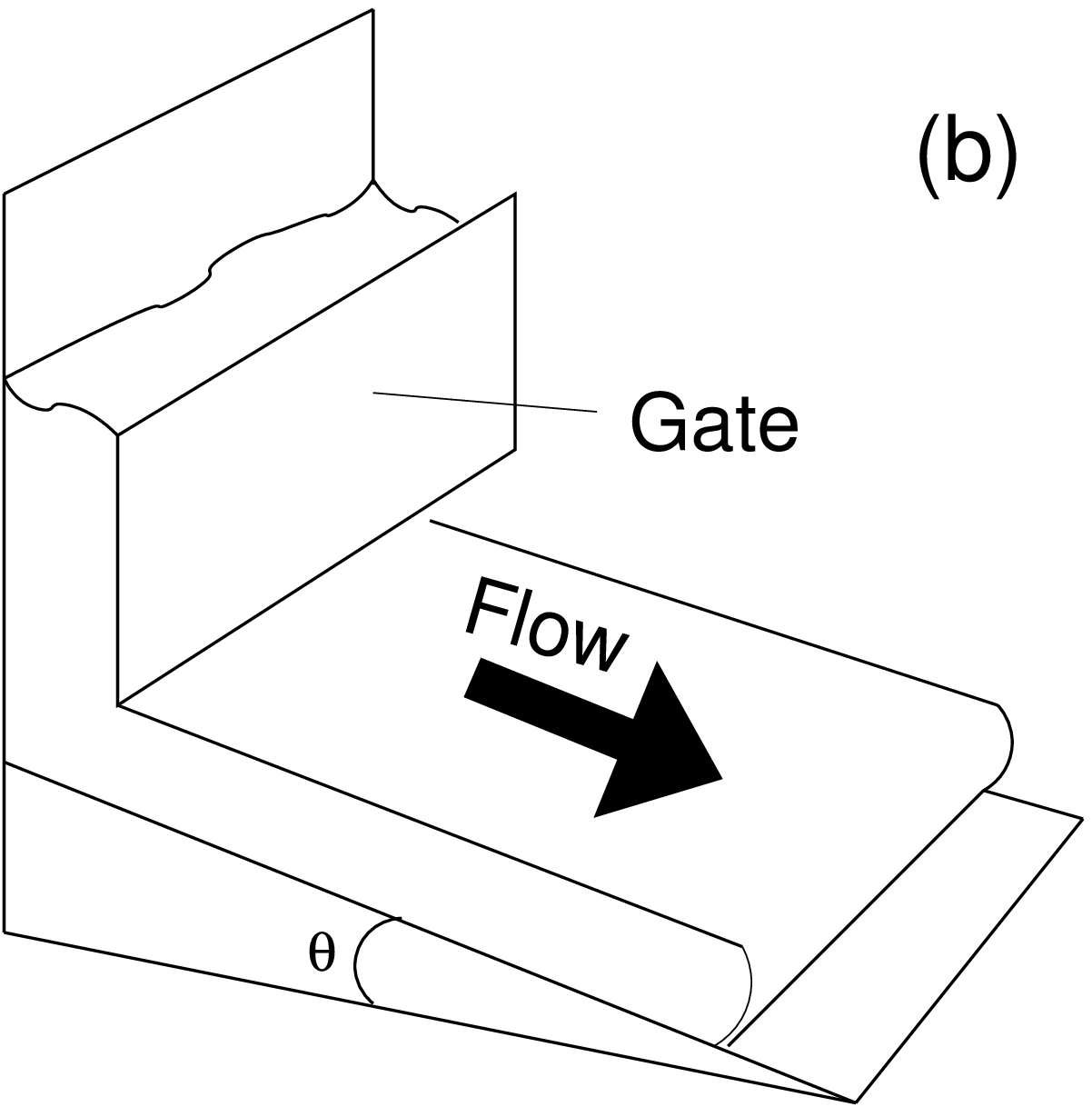}
\end{center} 
  \caption{Schematics of experiments used to study gravity-driven, granular
    flows. The (a) heap flow geometry used by Lemieux and Durian
    \cite{durian1} and Komatsu et al. \cite{komatsu1}, and (b) the
    inclined plane geometry employed by Pouliquen for chute flow studies
    \cite{pouliquen1}. Our simulations correspond closely to the
    Pouliquen experiments.}
  \label{chute3_fig1} 
\end{figure}

One method to study gravity-driven flows is to construct a
wedge-shaped static pile that evolves to produce an inclined surface
close to the angle of repose of the material
\cite{durian1,komatsu1,khakhar2}, Fig.~\ref{chute3_fig1}(a). We term
these {\it heap flows}. Here, the inclination of the surface is
determined by material properties. To study flow, material is supplied
via a hopper-feeder mechanism at the high end of the pile and flow is
induced at the free surface. In this case, although the flow rate may
be varied by changing the discharge rate of the hopper, the system
typically flows only in a thin surface layer, indicated by the shaded
region in Fig.~\ref{chute3_fig1}(a), which is of order 10 particles
diameters in depth. In one such experiment, Lemieux and Durian
\cite{durian1} observed avalanche behavior for low particle flux,
characterized by sporadic downstream movement at the free surface. On
increasing the particle flux, a velocity profile with depth was
observed to be approximately exponential in the thin surface layer.
More recently, utilizing a similar experimental geometry, Komatsu et
al.~\cite{komatsu1} found that the `stationary' bulk material beneath
the surface flow actually undergoes very slow, `creeping' motion with
a well-defined exponential velocity profile at longer times.

Another method uses inclined planes to study flows and avalanches,
typically with rough beds designed to reduce slip at the base, thereby
avoiding plug or unstable flow \cite{walton2,louge1,leo8}. We term
these {\it chute flows}. Roughness is introduced onto the chute by
gluing particles to the chute base \cite{pouliquen1,ancey2,aguirre1},
or by attaching a frictional material to the base, such as the velvet
cloth used by Daerr and Douady \cite{daerr1}. Chute flow studies of
two and three dimensional systems with rough bases report
depth-averaged velocities with both linear \cite{azanza1,ancey2} and
non-linear dependence \cite{pouliquen1,ancey2} on height. The geometry
for these experiments is shown in Fig.~\ref{chute3_fig1}(b).  The
material is contained in a reservoir at one end of a plane inclined at
a variable angle $\theta$ with respect to the horizontal. After the
gate to the reservoir is opened, material flows down the incline.
Steady flow occurs only at high enough angle; while flows come to rest
quickly on the rough bed for small inclination
angles~\cite{pouliquen2}.  Increasing the gate opening leads to
thicker flows down the incline: there are no dead zones along the
incline and all the material flows down the chute.  This allows one to
independently control the inclination angle and material flux of the
pile, in contrast to the heap experiments.

Using a range of inclination angles and flow depths, Pouliquen
\cite{pouliquen1} measured the depth-averaged flow velocity of
material for piles of different thickness. The velocity in the
direction of flow scaled with height of the pile as $v_{x}\propto
h^{3/2}$, which has a different form from the heap flow results
reported by Lemieux and Durian \cite{durian1} and Azanza et al.
\cite{azanza1}. However, Ancey's comprehensive experimental
investigation \cite{ancey2} demonstrated that during the course of one
experiment, a range of flow behaviors is obtained that depends, among
other parameters, on the flow rate of the material. Ancey's results
suggest that the different flow profiles observed are not necessarily
in contradiction, but may be a consequence of the conditions under
which the measurements are made.

In our earlier studies \cite{deniz2,leo7}, we simulated the rheology
of dense, cohesionless, frictional spheres in inclined plane
geometries. Employing periodic boundary conditions in the flow and
vorticity directions, our simulations can be considered equivalent to
the experimental set up in Fig.~\ref{chute3_fig1}(b), for an
infinitely long chute far away from the material reservoir and side
walls, with a continuous supply of material. Our previous study was
concerned with the flow of thick piles on rough bases over a range of
tilt angles where we found Bagnold scaling in stable steady flow, for
a range of parameters (e.g.  particle friction and inelasticity, and
inclination angles).

To investigate the discrepancies between the various experimental
results previously mentioned, we have extended our previous
computational studies of granular chute flows to cover a wide range of
pile thickness, from very thin to very thick piles, both near and far
from the angle of repose $\theta_{r}$. We show here that even thick
piles experience a crossover in behavior from Bagnold rheology to
linear velocity profiles, and also avalanche-like flow near the angle
of repose. We therefore refine our definition of stable, steady state
flow: we report on three distinct steady state flows observed in our
simulations which match the experimentally observed flows reported by
Lemieux and Durian \cite{durian1} (avalanches), Ancey \cite{ancey2}
(linear velocity profiles), and Pouliquen \cite{pouliquen1} (Bagnold
scaling).  Our main conclusion is that the observed behavior is
strongly dependent on how close the system is to jamming or how close
$\theta$ is to $\theta_{r}$.

In the next section we briefly describe the simulation procedure. In
section \ref{chute3_results} we present the results of our
simulations, in particular defining the phase space of our study and
describing transitions between the flow properties of thick and thin
piles both near and far from the angle of repose. We also present
results for our analyses on the initiation and cessation of flow in
section \ref{chute3_initiate}. In section \ref{chute3_conc} we
summarize the main conclusions of this work.

\section{Simulation Procedure}
Here we briefly introduce the simulation model; a thorough description
of the technique is available elsewhere \cite{deniz2,leo7}. We model
$N$ cohesionless, frictional, and inelastic spheres of diameter $d$
and mass $m$ flowing on a fixed, rough bed of area $A$ with a free top
surface.  Although most of our results are for monodisperse systems,
we have verified that our results hold for systems with small
polydispersity \cite{footnote12}.

We employed a 3D contact force model that captures the major features
of granular particles: interactions between spheres are modeled as
Hertzian contacts with {\it static friction}. The implementation of
the contact forces, both the normal forces and the shear (friction)
tangential forces, is essentially a reduced version of that employed
by Walton and Braun \cite{walton1}, developed earlier by Cundall and
Strack \cite{cundall1}. Normal and tangential forces acting on
particle $i$ due to contact with particle $j$ are given by
\begin{eqnarray}
{\bf F}_{n_{ij}}&=&\sqrt{\delta_{ij}/d}\left(k_{n}\delta_{ij}{\bf n}_{ij}
-m_{\rm eff}\gamma_{n}{\bf v}_{n_{ij}}\right), 
\label{chute3_eq3}\\
{\bf F}_{t_{ij}}&=&\sqrt{\delta_{ij}/d}\left(-k_{t}{\bf u}_{t_{ij}}
-m_{\rm eff}\gamma_{t}{\bf v}_{t_{ij}}\right), 
\label{chute3_eq4}
\end{eqnarray}
where $\delta_{ij}$ is the particle-particle overlap, ${\bf n}_{ij}
\equiv {\bf r}_{ij}/r_{ij}=({\bf r}_{i}-{\bf r}_{j})/|{\bf r}_{i}-{\bf
  r}_{j}|$, is the unit normal along the line of centers, for particle
$i$ situated at ${\bf r}_{i}$ and $j$ at ${\bf r}_{j}$. ${\bf
  v}_{n,t}$ are relative velocities in the normal and tangential
directions respectively, and ${\bf u}_{t_{ij}}$ is the elastic
tangential displacement that is set to zero at the initiation of a
contact.  $m_{\rm eff}$ is the effective mass and $k_{n,t}$ and
$\gamma_{n,t}$ are elastic and viscoelastic constants respectively.
Thus in a gravitational field {\bf g}, the force on particle $i$
arises from Newton's second law for the translational and rotational
degrees of freedom,
\begin{eqnarray} 
{\bf F}_{i}^{\rm tot}&=&m {\bf g} + \sum_{j}{{\bf F}_{n_{ij}}
+{\bf F}_{t_{ij}}}; \\
\mbox{\boldmath{$\tau$}}_{i}^{\rm tot}&=&-\frac{1}{2}\sum_{j}{{\bf r}_{ij}
\times{\bf F}_{t_{ij}}}, 
\label{chute3_eq6}
\end{eqnarray}
and the force on particle $j$ is determined from Newton's third law. 

\begin{figure}
\begin{center}
  \includegraphics[width=6cm]{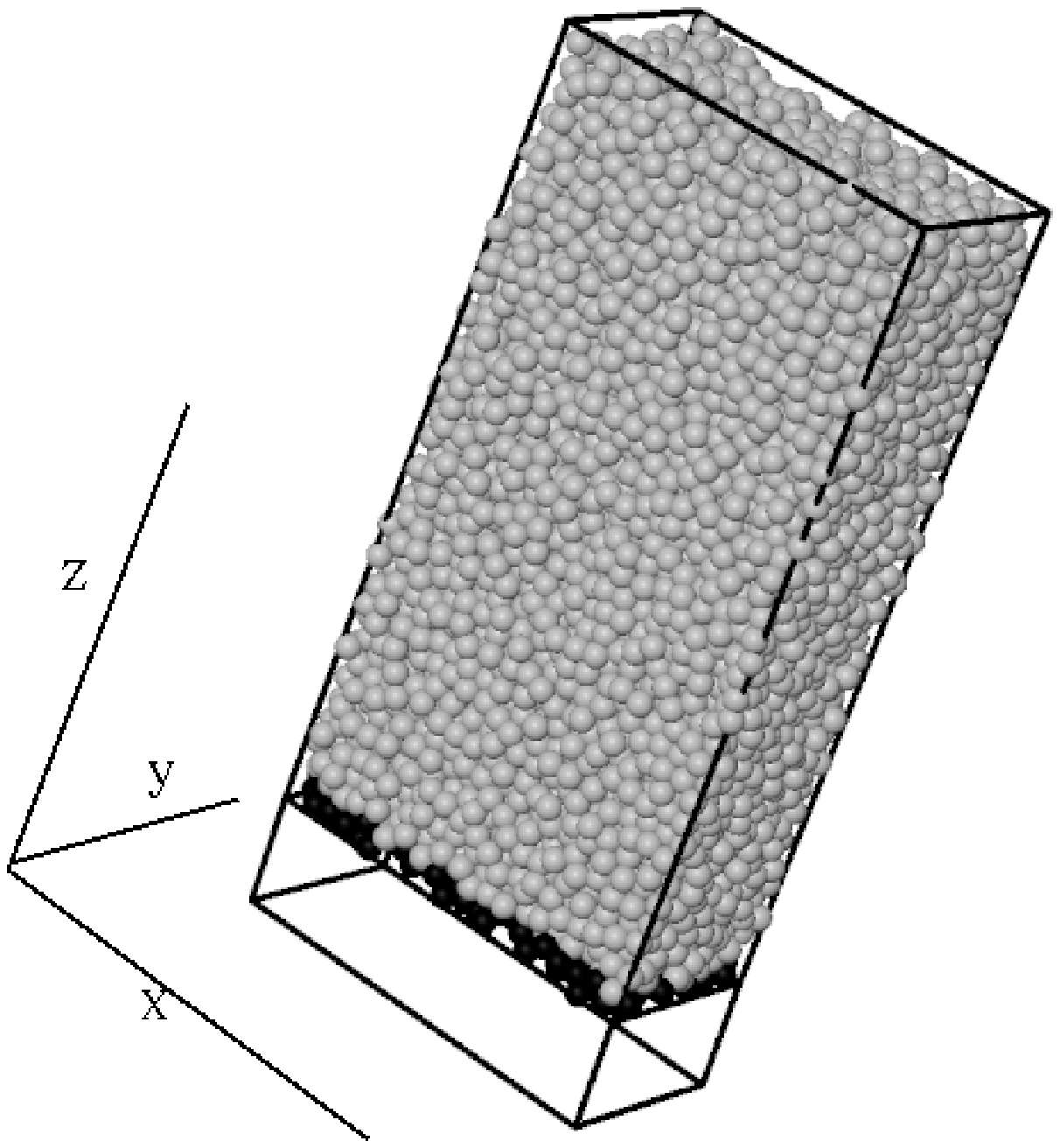}
  \includegraphics[width=6cm]{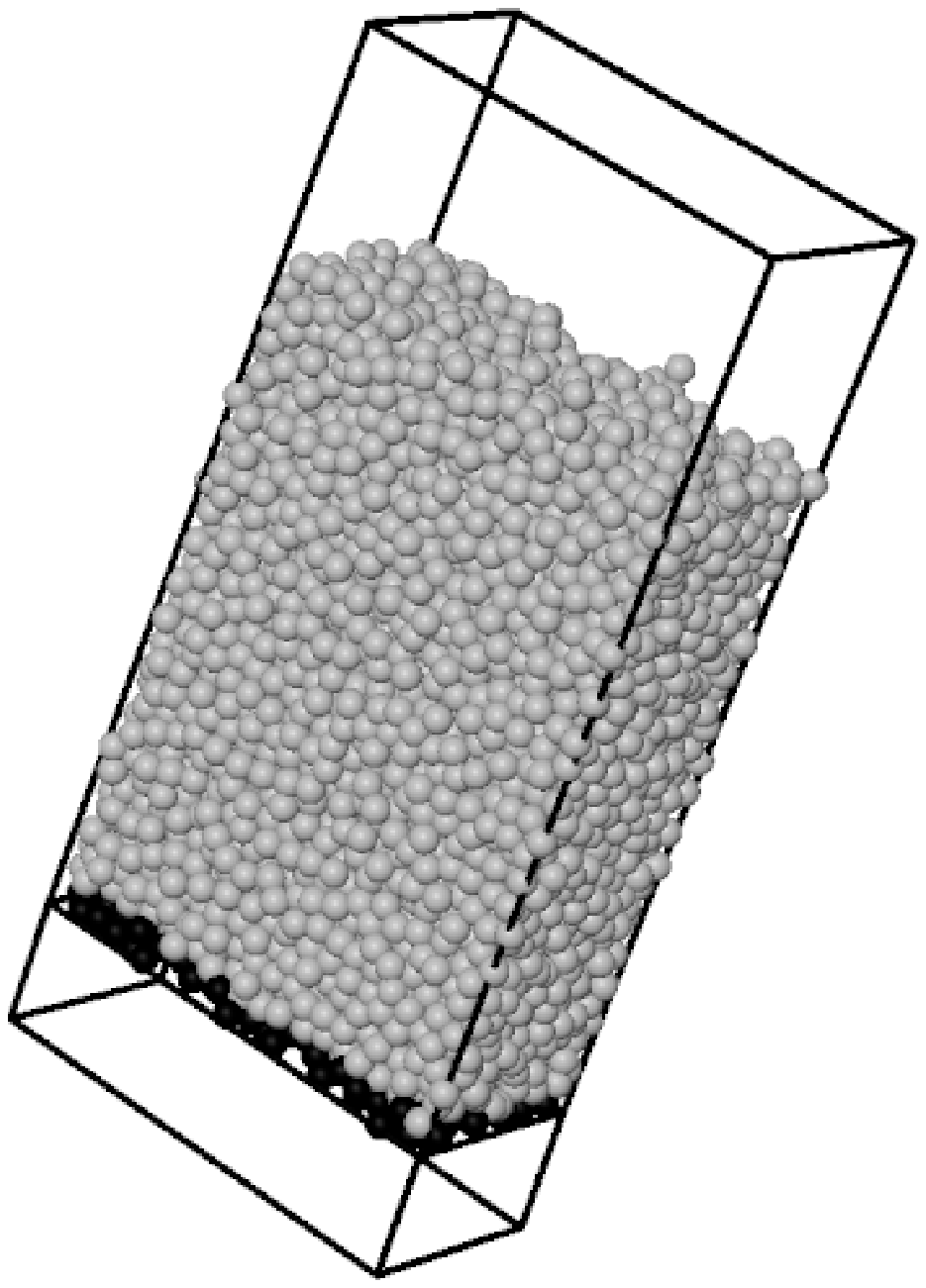}\\
(a)\hfil(b)\hfil\\
\bigskip
  \includegraphics[width=6cm]{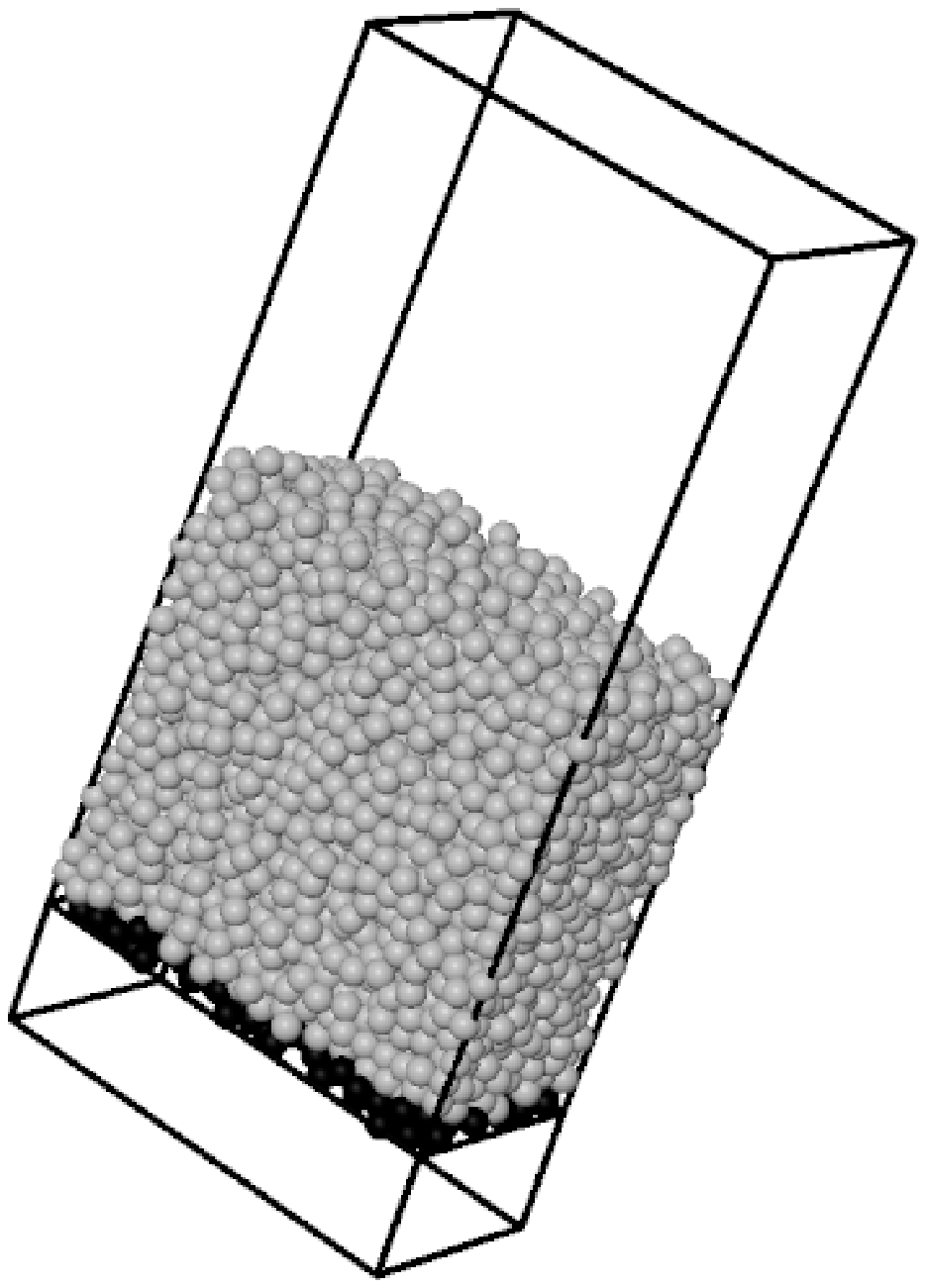}
  \includegraphics[width=6cm]{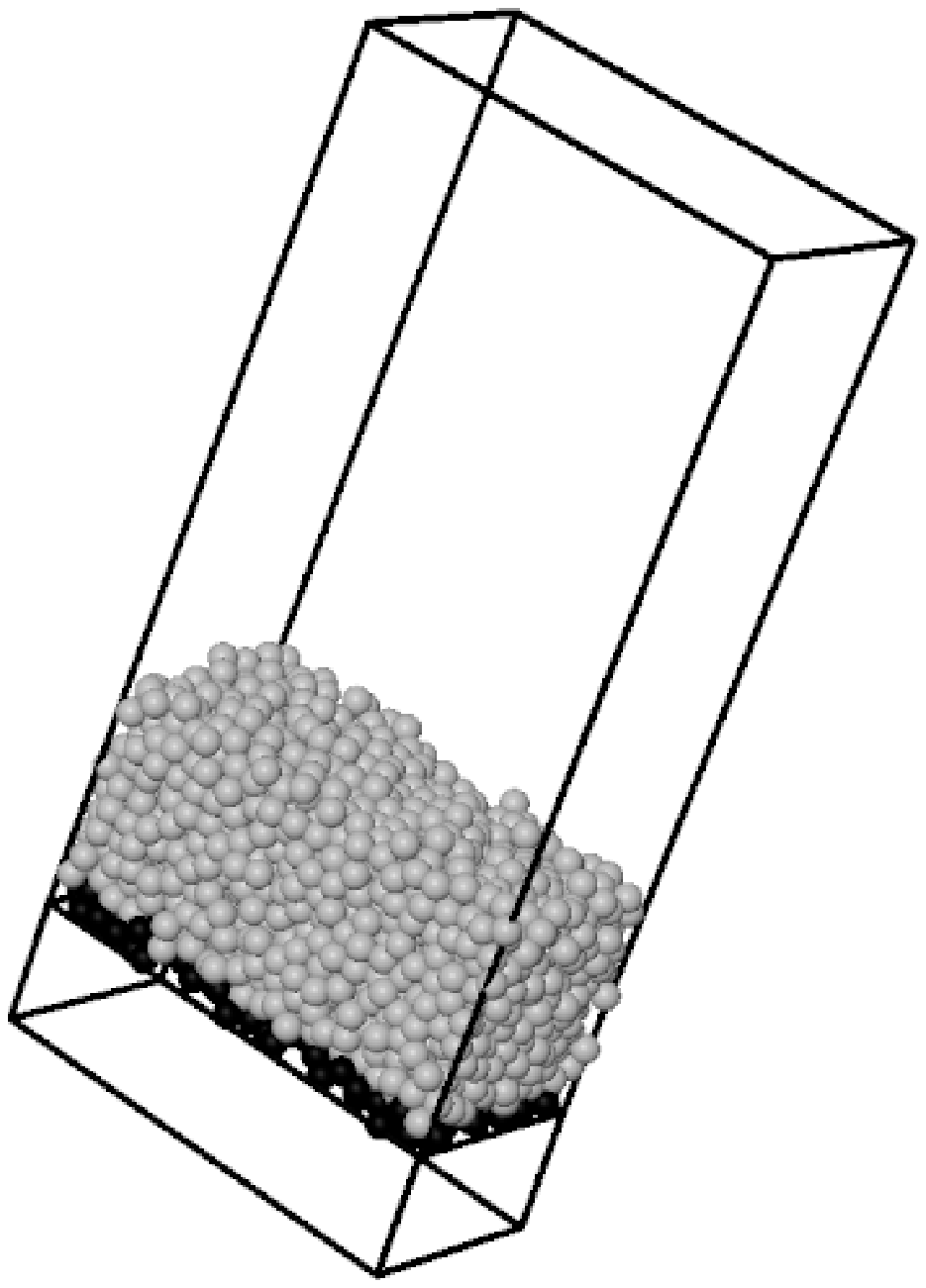}\\
(c)\hfil(d)\hfil\\
\end{center} 
  \caption{Configuration snapshots for systems of various sizes with a chute
    base of fixed area $A=200d^{2}$, that is inclined at an angle
    $\theta=24^{\circ}$ to the horizontal, where (a) $N=8000$, (b)
    $N=6000$, (c) $N=4000$, and (d) $N=2000$. The fixed rough base is
    denoted by the black particles, the flowing particles are grey, and
    the frame is a guide to the eye fixed at the height $h\approx 39.6d$
    of the $N=8000$ system to provide a comparison between the different
    flow heights.}
  \label{chute3_fig2} 
\end{figure}

As shown in Fig.~\ref{chute3_fig2}, we define flow to be parallel to
the $x$-direction, $z$ the height normal to the base, and $y$, the
vorticity axis, is normal to the $xz$-plane. Values for the friction
coefficient $\mu$, and the elastic and viscoelastic coefficients
(which are related to the moduli of the material) are chosen to be
characteristic of real materials. For all simulations in this study we
set $\mu=0.50$, $k_{n}=2\cdot10^{5}mg/d$, $k_{t}=\frac{2}{7}k_{n}$,
$\gamma_{n}=50\sqrt{g/d}$, and $\gamma_{t}=\gamma_{n}/2$
\cite{footnote17}. The time-step $\delta t=10^{-4}\tau$, where
$\tau=\sqrt{d/g}$. For a millimeter sized particle $\tau\approx0.01s$.
All simulations were run for at least $1\cdot10^{7}\delta t$ before
any quantities were measured.

Figure \ref{chute3_fig2} shows snapshots of flowing systems where the
area of the chute base $A\equiv L_{X}\times L_{Y}= 20d \times 10d =
200d^{2}$, and we vary the total number of particles $N$. The fixed
chute bed is made from a random close packing of particles with the
same diameter as those in the bulk, mimicking gluing particles onto a
chute base as in Ref.~\cite{pouliquen1}, for example. In section
\ref{depth} we kept $A=200d^{2}$ fixed. In section
\ref{chute3_initiate}, we used a base with dimensions $A = 200d\times
10d = 2000d^{2}$. We have shown \cite{leo7} that results for thick
piles are robust with respect to the dimensions of the chute base. The
height of a static pile $h_{\circ} = N/A\rho_{\circ}$ depends on the
density of the static packing $\rho_{\circ}$, or equivalently the
volume fraction $\phi_{\circ} \equiv \rho_{\circ}\pi/6~(\approx 0.60$
\cite{leo7}) which is defined in the hard sphere limit (where the
subscripts denote the static values). In the flowing state, the actual
height of the pile increases with inclination angle due to dilation
effects of the flow. However, the dependence of $h$, and therefore
$\rho$, on $\theta$ is weak. We find that the computed bulk volume
fractions vary over a small range, $0.55\lesssim\phi\lesssim 0.60$,
for steady state flows for the particular parameters chosen in this
paper \cite{leo7}.  Thus for the simulations $h_{\circ}=\pi
N/6A\phi_{\circ}$: for $N=5000$, $h_{\circ}\approx 22d$.

\section{Results}
\label{chute3_results}
In our previous studies \cite{deniz2,leo7}, we roughly determined the
range of parameters where steady state flow existed for chute flows,
focusing mainly on steady state flow for thick piles away from
$\theta_{r}$. We found that $\theta_{r}$ is approximately independent
of height for piles of sufficient depth. In this work we perform a
more thorough analysis of the entire region of the $h-\theta$ phase
space. We determine several different flow regimes within the stable,
steady state region which are qualitatively identical to several
experimental observations in inclined plane geometries, on avalanche
dynamics \cite{daerr1,daerr2} and rheology
\cite{pouliquen1,azanza1,ancey2,durian1}.  Although the exact location
of the boundaries of the phase picture shown in Fig.~\ref{chute3_fig3}
are dependent on specific particle parameters, such as the friction
and damping coefficients, the features we report here are general to
systems that exhibit steady state flow.

\subsection{Phase Behavior}
\label{chute3_res1}
The overall phase behaviour is determined by the height of the pile
$h$ and the angle of inclination $\theta$, as characterised in 2D by
Pouliquen and Renault \cite{pouliquen2}. A prominent feature of that
study is the existence of a {\it flow line}, later observed in 3D
experiments by Pouliquen \cite{pouliquen1} and also in studies of
avalanches \cite{daerr1}. The flow line defines the angle of repose as
a function of height of material flowing $\theta_{r}(h)$. One can
equivalently refer to this in terms of the conjugate variable
$h_{stop}(\theta)$ \cite{pouliquen1}: $\theta_{r}(h)$ is the angle at
which a pile of thickness $h$ will cease to flow, and conversely,
$h_{stop}(\theta)$ gives the maximum thickness of material that will
cease to flow at an inclination angle $\theta$. Various theoretical
treatments have also successfully predicted the existence of the flow
line \cite{aranson1,mills5}. Experimental \cite{aguirre1,daerr1} and
numerical \cite{barker7} studies have also identified distinct
sub-regions within the stable flow regime of which more will be
discussed below.

\begin{figure}
\begin{center}
  \includegraphics[width=8cm]{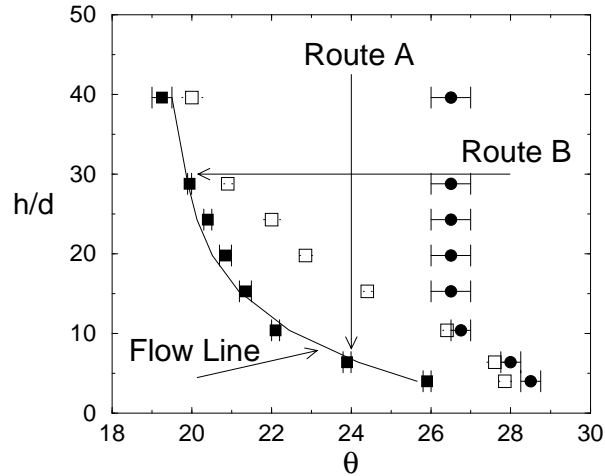}\\
\end{center} 
  \caption{The $h-\theta$ phase diagram for granular chute flow, where $h$ is
    the height of the pile. The filled squares are the simulation data
    for $\theta_{r}(h)$ and we indicate the {\it flow line} as a fit to
    Eq.~\ref{chute3_eq7}. The open squares are $\theta_{m}(h)$.  Solid
    circles are $\theta_{max}(h)$, the angle at which we observe the
    onset of unstable flow. Route A corresponds to a path taken by
    keeping the angle fixed and reducing the system size $N$, whereas
    Route B has $N$ fixed and decreasing inclination. These routes are
    discussed in section \ref{depth}.}
  \label{chute3_fig3} 
\end{figure}

In order to quantify this boundary with precision for our system, we
performed a number of simulations to generate the $h-\theta$ phase
diagram shown in Fig.~\ref{chute3_fig3}. Initially, a static packing
with a volume fraction $\phi\approx 0.60$, at $\theta=0^{\circ}$
\cite{leo9}, is tilted until flow is observed. The {\it maximum angle
  of stability} $\theta_{m}$ is defined as the angle at which flow is
initiated and largely depends on construction history of the initial
static packing, such as the roughness of the supporting bed and the
packing fraction of the initial configuration \cite{aguirre2}. The
angle of repose $\theta_{r}$ is defined as the inclination angle at
which the system subsequently jams and flow ceases upon reducing
$\theta$ back down.

Our initial packing was generated by taking a flowing system and
gradually reducing the inclination down to $0^{\circ}$ and then
allowing the system to relax until it became completely static.
Experimental studies on avalanches have shown that $\theta_{m}$ can
fluctuate over a range of $3^{\circ}-4^{\circ}$ for different packings
with the same height because of construction history effects
\cite{aguirre1}.

We have also observed avalanche-like dynamics in our simulations. We
characterise avalanches as distinct from regular or continuous flow by
observing the time evolution of bulk quantities of the system such as
bulk kinetic energy or stress. For continuous flow, the stress or bulk
energy will fluctuate about a well-defined mean value, with
fluctuations typically less than $10\%$ of the mean. In the case of
avalanches, the stress and energy fluctuate about a mean value with
O(1) fluctuations, indicative of sporadic flow that mainly occurs at
the free surface. We also checked any possible system size dependence.
To compare with our smaller systems, we studied a system with
$N=32000$ and $A=80d\times40d$ (i.e. approximately $10d$ high on a
base with an area four times our standard system size). We again
observed avalanche-like behaviour for the big system, although we
cannot rule out that in the limit of infinite system size, avalanches
might not be observed.

From our extensive study, we were able to determine the following
general features of flow down inclined planes: there are three
principal regions, corresponding to {\it no flow}, {\it stable flow},
and {\it unstable flow}. For a pile of given thickness $h$, and fixed
microscopic interaction parameters, these three regions are separated
by two angles: $\theta_{r}$, the angle of repose, and $\theta_{max}$,
the maximum angle for which we observe stable flow.

\begin{description}
\item[No flow] $\theta<\theta_{r}$: the system does not flow. Starting
  from $\theta=0^{\circ}$, an inclination of
  $\theta=\theta_{m}>\theta_{r}$ is required to initiate flow. As
  $\theta_{m}$ is approached, a few particles at the free surface
  rearrange and tumble short distances before coming to rest, but this
  does not lead to bulk motion or flow \cite{aguirre3}. $\theta_{r}$
  is the angle at which flow ceases on reducing the inclination from a
  flowing state - it is the lowest angle at which flow is observed and
  depends on material properties (friction and inelasticity)
  \cite{zhou1}, as well as the roughness of the supporting bed.
\item[Stable flow] $\theta_{r}<\theta<\theta_{max}$: bulk flow is
  observed (after flow is initiated). The bulk averaged kinetic energy
  of the system is constant with time and the entire pile is seen to
  be in motion while no slip is observed at the bottom with these
  rough boundary conditions. The stable flow region can be divided
  into the following three regimes:
\begin{description}
\item[Continuous flow I] $\theta>>\theta_{r}$: away from the flow
  line, the rheology is well approximated by Bagnold scaling,
  Eqs.~(\ref{chute3_eq1}) and (\ref{chute3_eq2}).
\item[Continuous flow II] $\theta_r <\theta\lesssim\theta_m$: close to
  the flow line, the rheology is not Bagnold-like, rather the velocity
  profile is better approximated by a linear relationship with depth
  $v(z)\propto z$. The upper boundary $\theta_{m}$ is also seen in
  experimental studies of avalanches and differentiates between
  different flow mechanisms, depending on whether one is above or
  below $\theta_{m}$ \cite{daerr1}.
\item[Avalanching flow] $\theta\approx\theta^{^{+}}_{r}$: Very close
  to the angle of repose, but just above it, regular continuous flow
  is not observed, rather in this regime, steady state involves slow,
  underlying creep motion of the entire pile combined with sporadic
  free surface motion down the incline. In the vicinity of
  $\theta_{r}$, extremely large energy fluctuations are observed
  \cite{leo10}.
\end{description}
\item[Unstable flow] $\theta>\theta_{max}$: the development of a shear
  thinning layer at the bottom of the pile results in lift-off and
  unstable acceleration of the entire pile. $\theta_{max}$ becomes
  approximately constant for $h\gtrsim10d$.
\end{description}

The flow descriptions above capture the principal observations of
several different experiments of gravity-driven flows
\cite{pouliquen1,ancey2,durian1,daerr1}. In particular, the transition
from no flow to avalanche flow at $\theta \approx \theta_{r}$ is
similar to the observations reported by Lemieux and Durian
\cite{durian1} for low material flux. The transition from avalanches
to our continuous flow II regime appears to resemble the observations
reported by Daerr and Douady \cite{daerr1} of avalanche studies for
$\theta_{r}<\theta<\theta_{m}$. Similar observations have been
reported for rotating cylinder experiments \cite{rajchenbach1}.

Pouliquen \cite{pouliquen1}, later verified by Daerr and Douady
\cite{daerr1}, characterized the flow line - the dividing line between
no flow and flow. Using an empirical argument, Pouliquen suggested the
following relation:
\begin{equation}
\tan\theta_{r}=\tan\theta_{1}+[\tan\theta_{2}-\tan\theta_{1}]\exp\left(-\frac{h}{l}\right),
\label{chute3_eq7}
\end{equation}
where $\theta_{1}$ is the angle at which $\theta_{r}$ becomes
independent of depth, $\theta_{2}$ is the largest value of
$\theta_{r}$, and $l$ is a characteristic length scale. In
Fig.~\ref{chute3_fig3} we fit Eq.~\ref{chute3_eq7} to our simulation
points and obtained the following fitting parameters;
$\theta_{1}=19.60^{\circ}~(19.55^{\circ})$,
$\theta_{2}=28.50^{\circ}~(27.75^{\circ})$, and $l\approx8.5d$, where
the values in parentheses are the data points from the simulations.
Our simulations not only predict the phase behavior of chute flows,
but also capture the subtle features emerging in recent experiments.
In the following sections we present various depth profiles of
density, velocity, and stress to better quantify the transitions
between thin and thick piles.

\subsection{Depth Profiles}
\label{depth}
We computed depth profiles as a function of $z$ for the packing
fraction $\phi$, velocity in the direction of flow $v_{x}$,
strain-rate $\dot{\gamma}$, shear stress $\sigma_{xz}$, and granular
temperature $T$ (see definition below), for a range of system sizes
and inclination angles. Since our simulations employ periodic boundary
conditions in the plane perpendicular to height, our simulations are
not influenced by the effects of side walls and are therefore
equivalent to measurements made along the centre line of a real chute
flow, away from the edges. It has been recognised that profiles taken
at the side walls are significantly different from those made along
the centre line \cite{ancey2,walton3}.

Because of the dependence of $\theta_{r}$ on $h$, we approached
$\theta_{r}$ in two ways: fixing the inclination angle of the chute
and observing flow in systems of different heights until flow ceased
for a pile of height $h_{stop}(\theta)$ (route A). Alternatively, we
held $N$ fixed and reduced the inclination angle of a flowing state
down to $\theta_{r}(h)$ (route B).

\begin{figure}
\begin{center}
  \includegraphics[width=6.5cm]{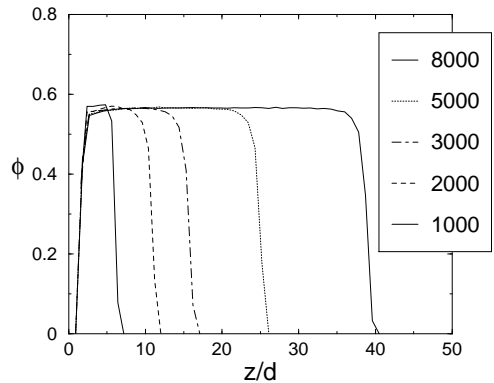}
\end{center} 
  \caption{Depth profiles of the packing fraction $\phi$, for systems of various
    size $N$, varying in heights from $h\approx 40d$ to $5d$ for
    $\theta=24^{\circ}$: route A in Fig.~\ref{chute3_fig3}.}
  \label{chute3_fig7} 
\end{figure}

\begin{figure}
\begin{center}
  \includegraphics[width=6.5cm]{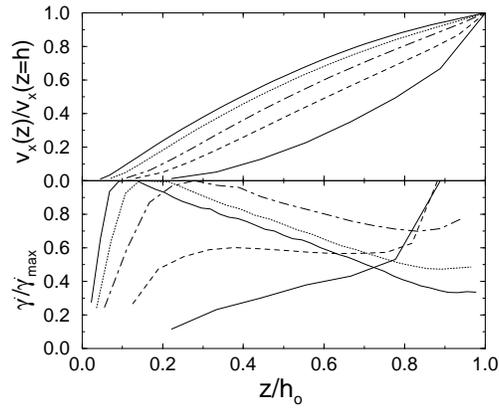}
\end{center} 
  \caption{Depth profiles of the velocity in the direction of flow (top panel)
    and the corresponding strain-rate (bottom panel), each normalized by
    their maximum value, for various system sizes for
    $\theta=24^{\circ}$. Note the depth is normalized by the respective
    height for clarity. Legend as in Fig.~\ref{chute3_fig7}.}
  \label{chute3_fig8} 
\end{figure}

\begin{figure}
\begin{center}
  \includegraphics[width=6.5cm]{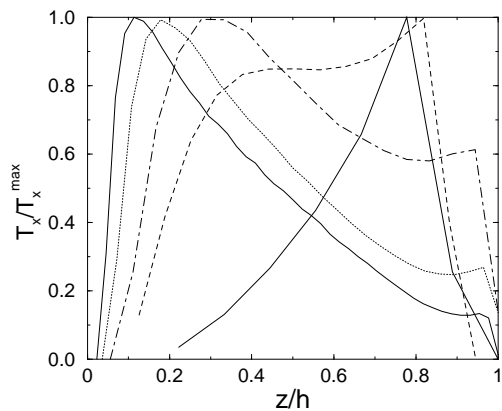}
\end{center} 
  \caption{Profile of the granular temperature $T_{x}/T_{x}^{max}$, in the
    direction of flow as a function of $z$ for various system sizes for
    $\theta=24^{\circ}$. Legend as in Fig.~\ref{chute3_fig7}.}
  \label{chute3_fig9} 
\end{figure}

\begin{figure}
\begin{center}
  \includegraphics[width=6.5cm]{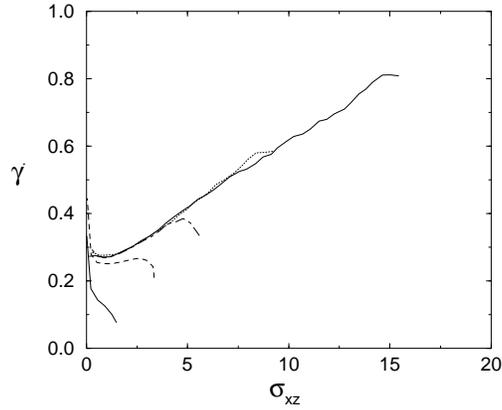}
\end{center} 
  \caption{Rheology data for various system sizes for $\theta=24^{\circ}$:
    $\dot{\gamma}$ vs.~$\sigma_{xz}$. Legend as in
    Fig.~\ref{chute3_fig7}.}
  \label{chute3_fig10} 
\end{figure}

{\bf Route A:} Figures \ref{chute3_fig7}-\ref{chute3_fig10} show
various bulk profiles for route A, that is, systems at a fixed
inclination angle of $\theta=24^{\circ}$ but of varying sizes,
$N=8000,~5000,~3000,~2000,~1000$. The density profiles in
Fig.~\ref{chute3_fig7} exhibit a constant density region throughout
the bulk of the material which is a familiar observation for thicker
systems \cite{leo7}.

The flow velocity $v_{x}$ and strain-rate $\dot{\gamma}$ profiles
shown in Fig.~\ref{chute3_fig8} are for the same systems reported in
Fig.~\ref{chute3_fig7}. For moderate to thick piles, $h\gtrsim 15d$,
the velocities and corresponding strain-rates scale with height.
However, for $h\lesssim 15d$, there are significant differences. The
velocity profiles, top panel in Fig.~\ref{chute3_fig8}, for the
thinner piles at first become more linear with depth, then ultimately
have a curvature opposite to that of the thicker piles. This gradient
transformation is more apparent from the strain-rate curves (bottom
panel Fig.~\ref{chute3_fig8}): for thicker piles $\dot{\gamma}$
decreases from the chute base toward the free surface of the flowing
pile, whereas in thinner piles there is a range of system sizes where
$\dot{\gamma}\approx constant$ throughout the pile, leading to a
linear velocity profile. For thinner systems still, $\dot{\gamma}$
increases near the free surface.

In this work, we define the ``granular temperature'' $T$ to be the
averaged velocity fluctuations about the mean bulk flow velocity at
that height, $T\equiv\langle v^{2}\rangle-\langle v\rangle^{2}$. The
temperature in the flow direction is defined as $T_{x}=\langle
v_{x}^{2}\rangle-\langle v_{x}\rangle^{2}$. Figure \ref{chute3_fig9}
shows the change in the profiles of $T_{x}$ from thick to thin piles,
which is similar to the behavior of $\dot{\gamma}$. From the
simulations we find that for {\it all sizes},
\begin{equation}
T_{x}\propto \dot{\gamma}^{2},
\label{chute3_eq8}
\end{equation}
although the data for thinner piles is considerably noisier.

Similarly, the rheology plot in Fig.~\ref{chute3_fig10} shows that
thicker piles display Bagnold rheology: $\sigma_{xz}\propto
\dot{\gamma}^{2}$, but there is a transition away from this as the
piles become thinner. For thin piles, $5\lesssim\frac{h}{d}\lesssim
15$, the strain rate is approximately constant away from the free
surface and the chute base.  For the thinnest piles $h\approx5d$, the
strain rate decreases with shear stress. This may come about from
competing length scales in the system, i.e. the thickness of the flow
determines the separation between the top free surface and the chute
base. Some theoretical treatments based on these considerations have
been used to explain the differences between observed velocity
profiles in flowing piles \cite{mills4}.

\begin{figure}
\begin{center}
  \includegraphics[width=6.5cm]{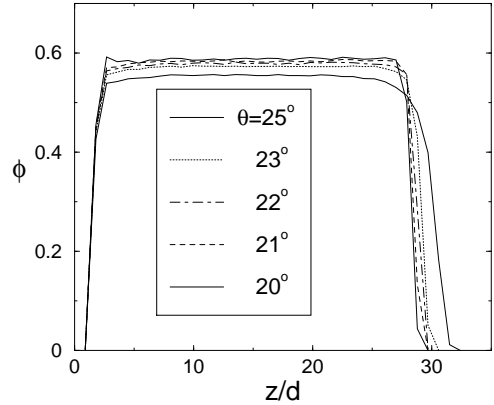}
\end{center} 
  \caption{Depth profiles of the packing fraction $\phi$, for a system with a
    fixed number of particles, $N=6000$, over a range of inclination
    angles, route B in Fig.~\ref{chute3_fig3}.}
  \label{chute3_fig11} 
\end{figure}

\begin{figure}
\begin{center}
  \includegraphics[width=6.5cm]{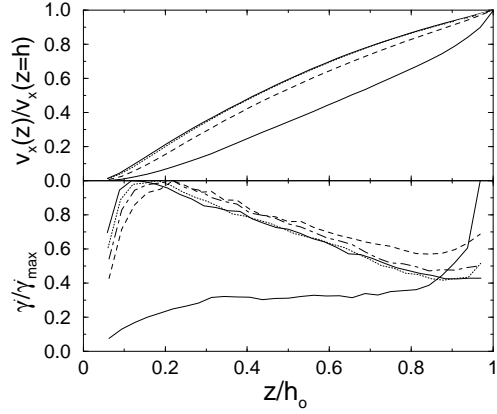}
\end{center} 
  \caption{Profiles for the velocity in the direction of flow $v_{x}$ (top
    panel) and the corresponding strain-rate $\dot{\gamma}$ (bottom
    panel), each normalized by their respective maximum values, for a
    system with a fixed number of particles, $N=6000$, over a range of
    inclination angles inclination angle. The height is normalized by
    the system size. Legend as in Fig.~\ref{chute3_fig11}.}
  \label{chute3_fig12} 
\end{figure}

\begin{figure}
\begin{center}
  \includegraphics[width=6.5cm]{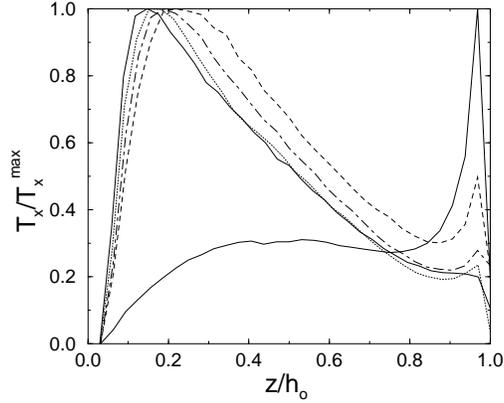}
\end{center} 
  \caption{Profile of the x-component of the granular temperature
    $T_{x}/T_{x}^{max}$ for a system with a fixed number of particles,
    $N=6000$, over a range of inclination angles. Legend as in
    Fig.~\ref{chute3_fig11}.}
  \label{chute3_fig13} 
\end{figure}

\begin{figure}
\begin{center}
  \includegraphics[width=6.5cm]{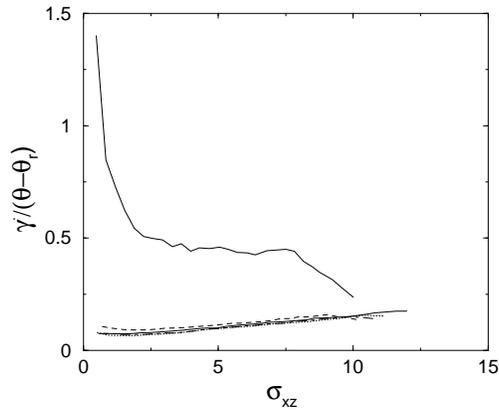}
\end{center} 
  \caption{Rheology data for a system with a fixed number of particles $N=6000$,
    over a range of inclination angles: $\dot{\gamma}$ normalized by
    $(\theta - \theta_{r})$ vs.~$\sigma_{xz}$. Legend as in
    Fig.~\ref{chute3_fig11}.}
  \label{chute3_fig14} 
\end{figure}

{\bf Route B:} Figures \ref{chute3_fig11}-\ref{chute3_fig14} show
depth profiles for a system taken along route B in
Fig.~\ref{chute3_fig3}. These are for a system of fixed size $N=6000$,
where we vary the inclination angle down toward $\theta_{r}$. In
Fig.~\ref{chute3_fig11} the density profiles for the system $N=6000$
are shown for a range of inclination angles
$\theta=25^{\circ},~23^{\circ}~,22^{\circ},~21^{\circ}~,20^{\circ}$.
We see the constant density region, where the bulk density decreases
with increasing angle due to dilatancy effects. However, there are no
qualitative differences between the behaviour at different angles.

For the velocity and strain-rate profiles shown in
Fig.~\ref{chute3_fig12}, differences do emerge, and these differences
mimic the behavior observed along route A. Away from the flow line,
the velocity profiles scale as Eq.~\ref{chute3_eq2}, whereas for
systems at lower angles the velocity profiles change to linear.
$\dot{\gamma}$ changes likewise as does the granular temperature
(shown in Fig.~\ref{chute3_fig13}).

In Fig.~\ref{chute3_fig14} we plot the rheology curves for the
different angles. We have rescaled $\dot{\gamma}$ by
$(\theta-\theta_{r})$, for convenience. We find that for systems with
$(\theta-\theta_{r})\gtrsim 1.0^{\circ}$, the rheology plots scale
well. However, for the system that is closest to $\theta_{r}$, the
flow behavior is very different, though similar to the differences
observed along Route A.

Therefore the two routes A and B indicated in Fig.~\ref{chute3_fig3}
offer complementary paths to similar states. Because of the very
nature of the dependence of $\theta_{r}$ on $h$, the flow properties
of systems at different angles and different heights show similarities
provided they are taken equivalently close to $\theta_{r}$.

\subsection{Velocity Scaling}
Pouliquen \cite{pouliquen1} showed that flows of thick and thin piles
approximately obey a simple scaling relation. Non-dimensionalizing
$v$, the velocity measured along the centre line at the top of the
flowing pile by $\sqrt{gh}$, where g is the acceleration due to
gravity and $h$ is the measured height of the pile, one obtains the
{\it Froude} number $Fr\equiv v/\sqrt{gh}$. By plotting the Froude
number against $h/h_{stop}$, where $h_{stop}(\theta)$ is the height of
a pile at which a system flowing at an angle $\theta$ would stop,
Pouliquen was able to collapse his full data set of systems with
different particle size and properties. Pouliquen proposed the simple
scaling relation,
\begin{equation}
\frac{v}{\sqrt{gh}} = \beta \frac{h}{h_{stop}}.
\label{chute3_eq9}
\end{equation}
where the coefficient $\beta$ was fit to the experimental data,
resulting in an apparently universal value, $\beta_{exp}=0.136$.

\begin{figure}
\begin{center}
  \includegraphics[width=8cm]{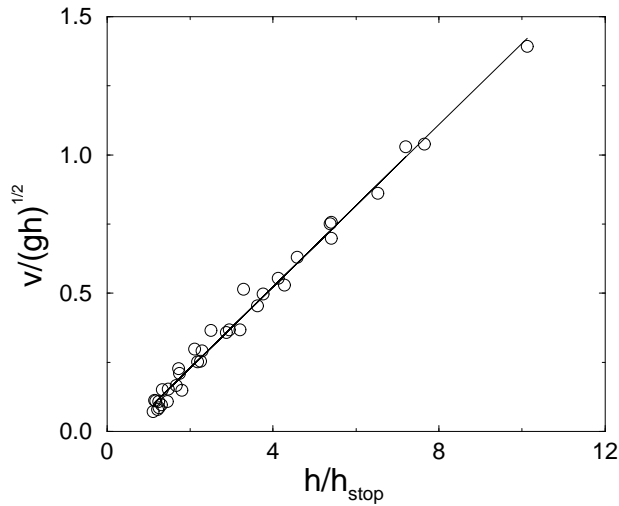}
\end{center} 
  \caption{Froude number $v/\sqrt{gh}$ as a function of $h/h_{stop}$ for
    different inclinations. A measure of the slope gives $\beta_{sim} =
    0.147$, compared with Pouliquen's experimental determination of
    $\beta_{exp} = 0.136$.}
  \label{chute3_fig15} 
\end{figure}

The quantity $h_{stop}$ encodes details of the particle roughness, the
base roughness and all other intrinsic properties of a particular
system. Accordingly, one need only measure $h_{stop}$ of the system in
question to determine the flow properties. In a similar vein, in
Fig.~\ref{chute3_fig15} we plot the flow velocity, taken from the top
of the flowing pile, non-dimensionalised by $\sqrt{gh}$ for each
height $h$ against $h/h_{stop}$. The data agrees well with a linear
relation, giving a coefficient $\beta_{sim}=0.147$, which is in good
agreement with the experimental result determined by Pouliquen
\cite{pouliquen1}, $\beta_{exp}=0.136$.

The scaling relation Eq.~\ref{chute3_eq9} although simple, appears to
be quite robust. Pouliquen \cite{pouliquen4} and also Andreotti et al.
\cite{daerr5}, have taken Eqs.~\ref{chute3_eq9} and \ref{chute3_eq7}
to predict phase behaviour and flow properties.  Their resulting
analyses of granular fronts captures many of the main features of
granular flows. However, as yet, there is no underlying explanation
for these seemingly simple relationships.

\section{Initiation and Cessation of flow}
\label{chute3_initiate}
We have also studied the initiation and cessation of flow, something
difficult to probe experimentally. For these studies we used large
systems, $A = 200d \times 10d = 2000d^2$ and $h_{\circ} \approx 50d$,
providing better resolution for these processes. In addition, any size
effects due to thin samples do not complicate our observations.

We have previously studied initiation of flow from a crystalline
static state in 2D \cite{leo7}. After such a system is tilted, a
disordering front propagates upwards from the bottom of the pile,
melting the ordered structure, until the entire pile is disordered and
flows in its steady state form.  Although the starting state in $3D$
need not be ordered, we find that flow initiation in $3D$ also
undergoes a similar bottom-up process.

\begin{figure}
\begin{center}
  \includegraphics[width=6.5cm,angle=270]{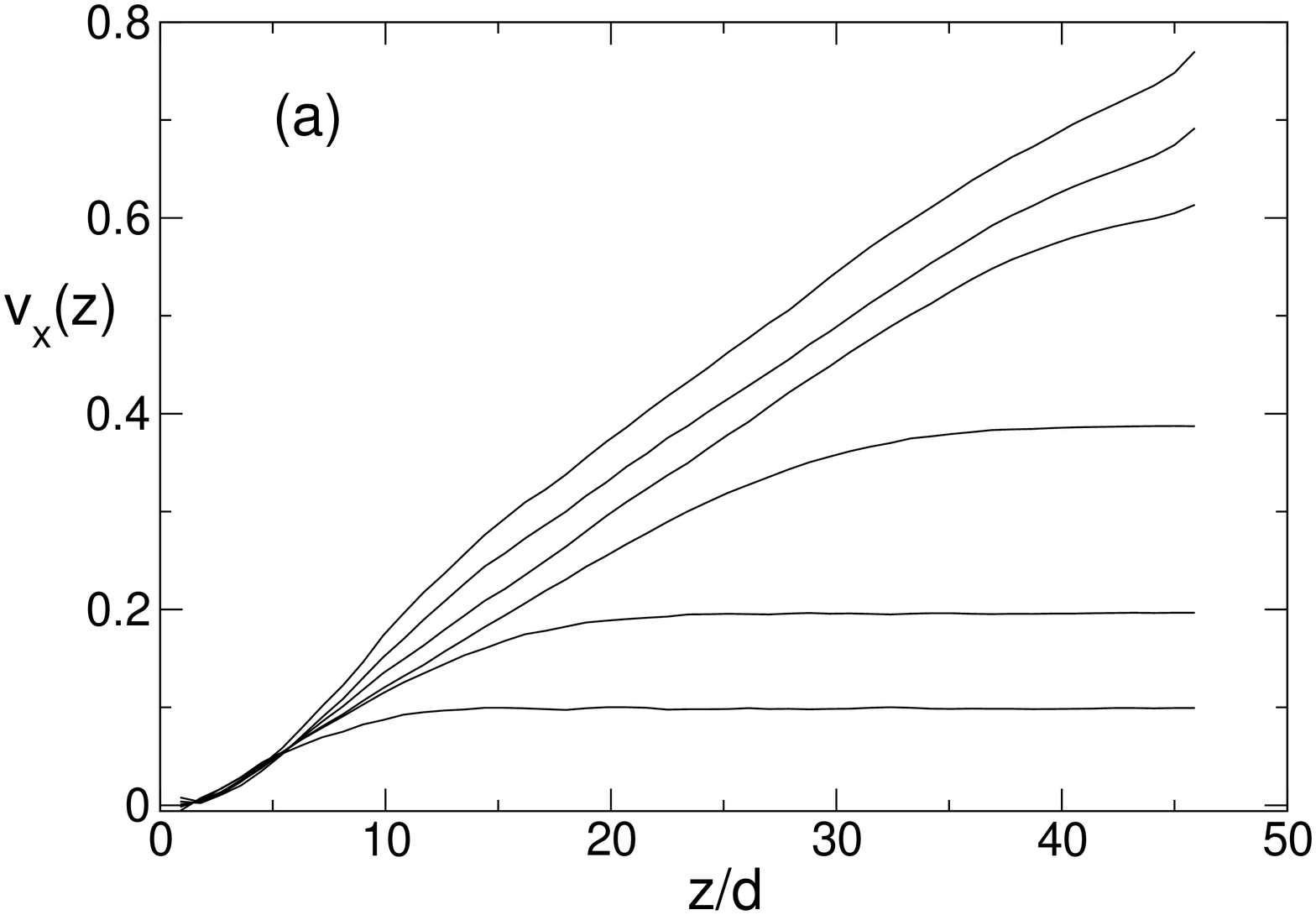} 
  \includegraphics[width=6.5cm,angle=270]{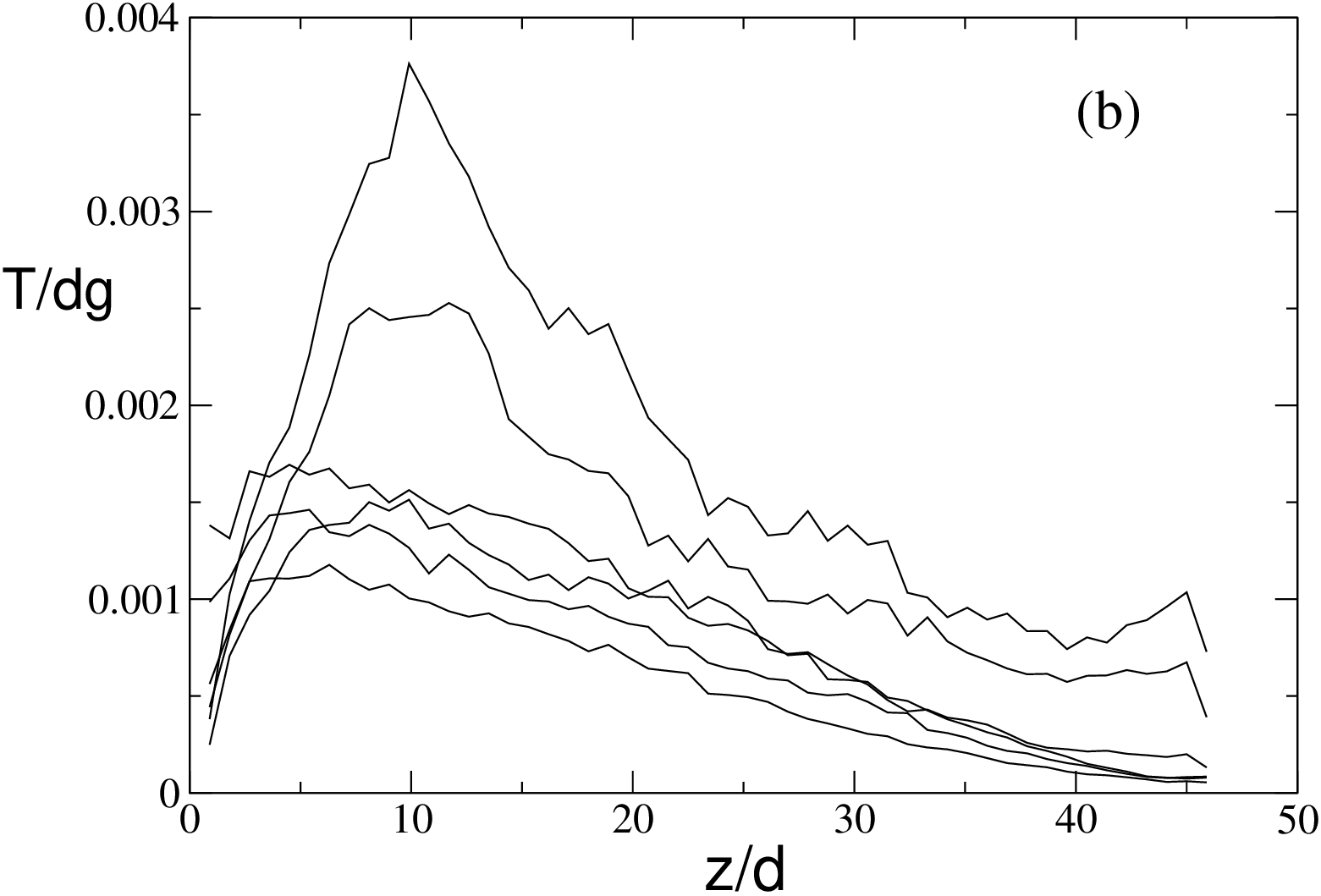}\\
\end{center} 
  \caption{Initiation of flow: time evolution of the depth profiles of, (a)
    the velocity in the direction of flow and (b) the granular
    temperature in the direction of flow, at the initiation of flow
    for a static state ($h\approx 50d$), initially at
    $\theta=0^{\circ}$, that is tilted to $24^{\circ}$. Initial
    failure occurs near the base, the pile then ``heats up'' from the
    bottom of the pile, a thermalization front propagates up the pile
    with time, disrupting the plug flow above it. The lowest curve
    starts at $t=025\tau$ after the system is first tilted, and we
    plot curves for $t=0.50\tau,~1\tau,~2\tau,~3\tau$, and $4\tau$
    respectively.}
  \label{chute3_fig5} 
\end{figure}

{\bf Initiation} was studied for a disordered, static packing of our
larger system that was tilted from $\theta=0^{\circ}$ to $24^{\circ}$.
See Fig.~\ref{chute3_fig5}. At the instant of flow initiation, the
system slips at the bottom boundary and begins to flow as a plug as
indicated in Fig.~\ref{chute3_fig5}(a). Particles closest to the base
undergo collisional motion with the fixed bed leading to the creation
of a granular temperature front, as shown in
Fig.~\ref{chute3_fig5}(b). This {\it thermalization} front propagates
up the pile, disrupting the plug region. After the original initiation
slippage, the slip region disappears and the velocity profile appears
approximately linear with height, with a crossover to the plug region.
After the plug phase melts, the system dilates slightly, leading to a
velocity profile of the Bagnold steady state form.

\begin{figure}
\begin{center}
  \includegraphics[width=6.5cm,angle=270]{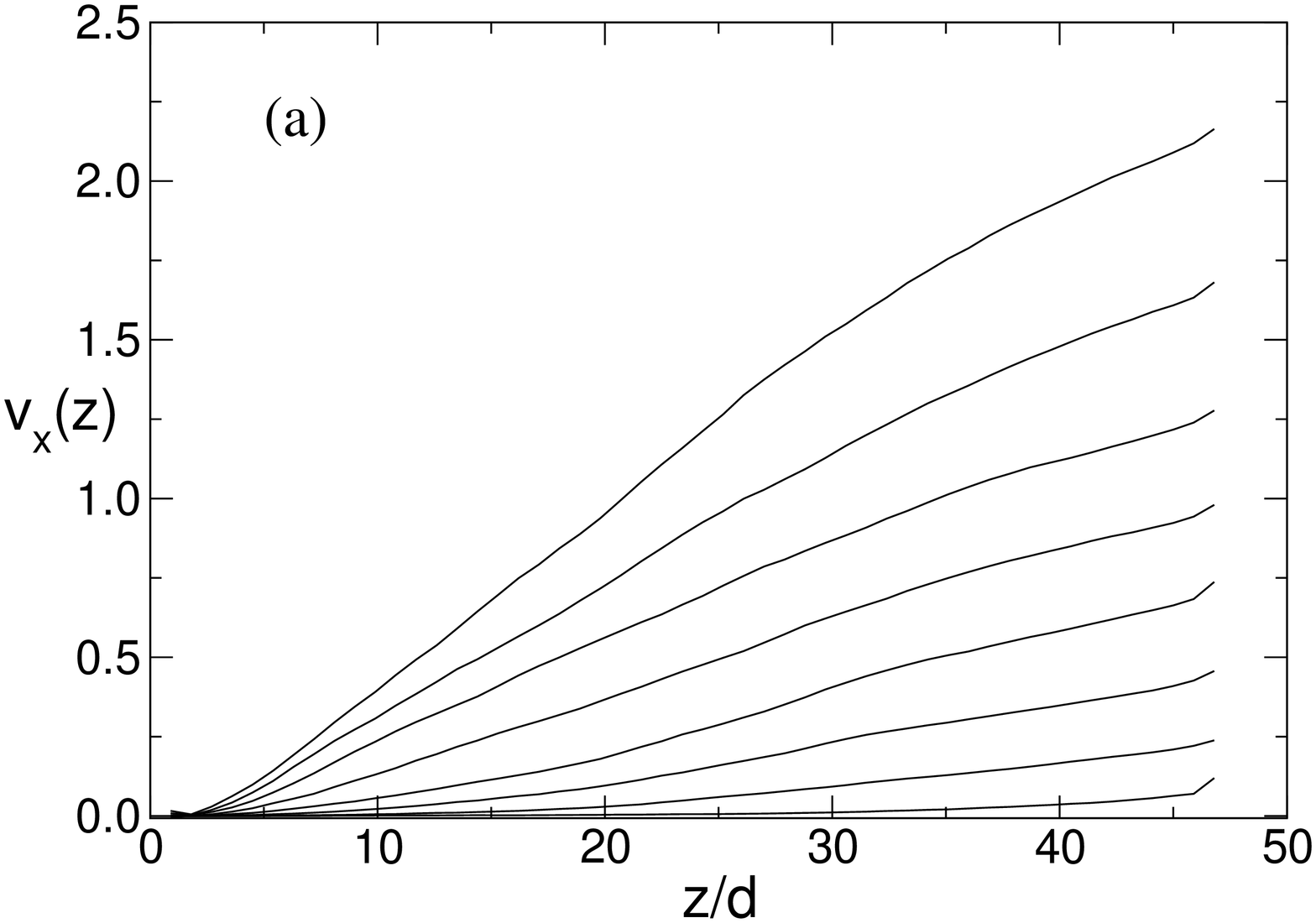}
  \includegraphics[width=6.5cm,angle=270]{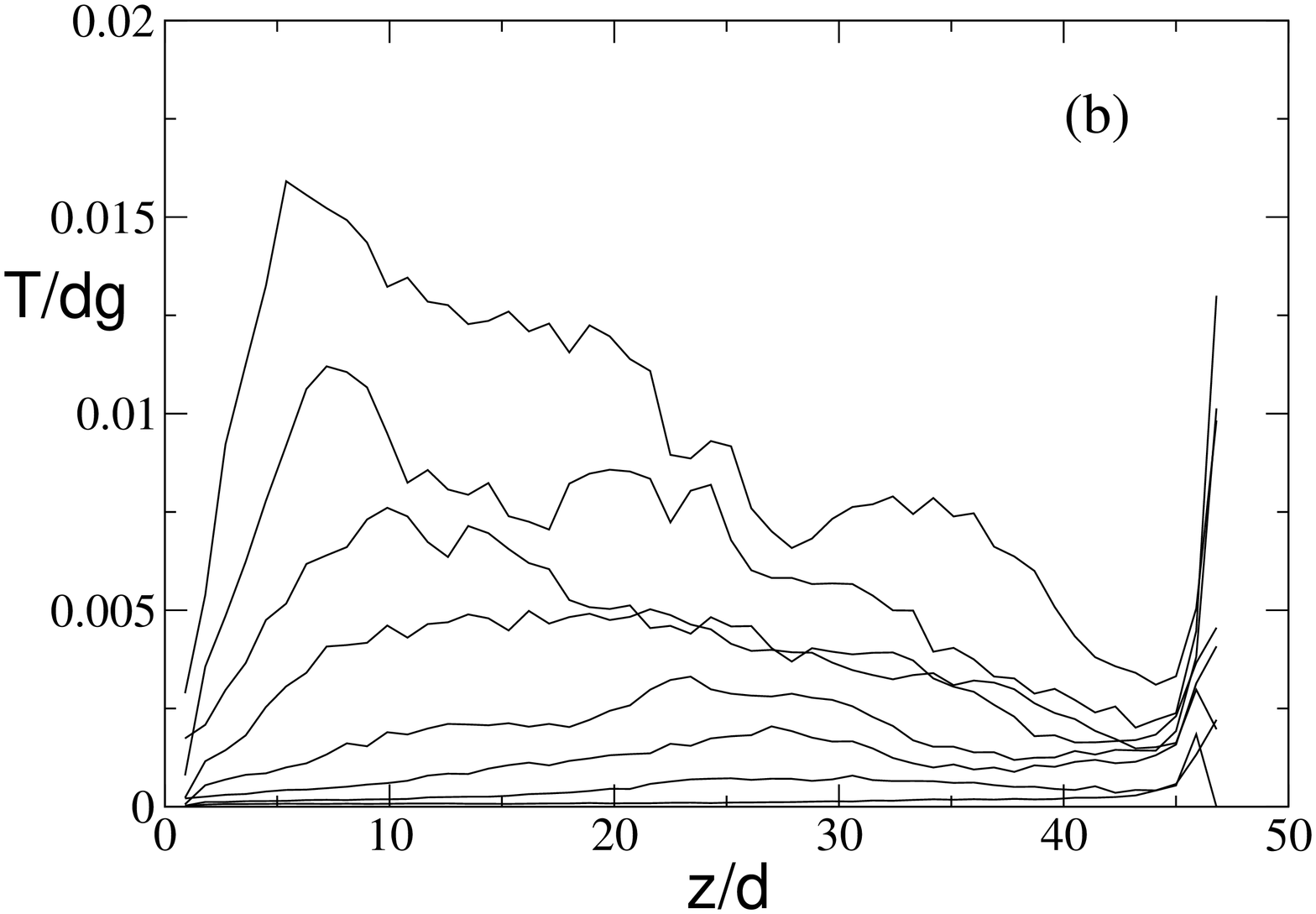}\\
\end{center} 
  \caption{Cessation of flow: time evolution of the depth profiles of, (a) the
    velocity in the direction of flow and (b) the granular temperature
    in the direction of flow, for the cessation of flow of a system
    with $h\approx 50d$.  Initially in steady state flow with
    $\theta=24^{\circ}$, the inclination angle is then reduced to
    $18^{\circ}<\theta_{r}$ leading to the gradual stopping of flow
    into a jammed state. Particles moving slowest at the base stop
    flowing first, leading to a freezing front that propagates up the
    pile. The top curve is taken at a time $100\tau$ after the angle
    is first reduced to below the angle of repose, and subsequent
    curves are separated by $10\tau$ later.}
  \label{chute3_fig6} 
\end{figure}

{\bf Cessation} was observed for a system that was initially in steady
state flow at $\theta = 24^{\circ}$, as the inclination angle was
reduced to $\theta = 18^{\circ}$, which is below $\theta_{r}$. See
Fig.~\ref{chute3_fig6}. Although it may seem intuitive to expect an
exact reversal of the initiation process, this is not observed.  After
$\theta$ is reduced to below the angle of repose, particles at the
bottom of the pile, which have the smallest velocities, come to rest
first becoming trapped by the rough base. The frozen layer then serves
to trap the slowly moving particles above it.  This {\it trapping}
front propagates up the pile towards the free surface, as indicated in
Fig.~\ref{chute3_fig6}(a). A consequence of this is that the granular
temperature cools from the chute base as the whole system gradually
comes to rest. We note the emergence of a thin saltating region at the
free surface during the cooling process.

During this trapping process the pile exhibits two distinct behaviors
separated by the trapping front. Below the trapping front, the
particles are frozen into a pack, whereas above the trapping front,
particles continue to flow. As the trapping front moves up the pile,
the depth of the flowing section continuously shrinks with time. As
shown in Fig.~\ref{chute3_fig6}(a), this leads to a bulk velocity
profile that consequently transforms from Bagnold-like flow to linear
with depth until flow ceases. Essentially, flow cessation is a
transient version of route B in Fig.~\ref{chute3_fig3} for flow at one
angle over a range of flow heights.

\begin{figure}
\begin{center}
  \includegraphics[width=8cm]{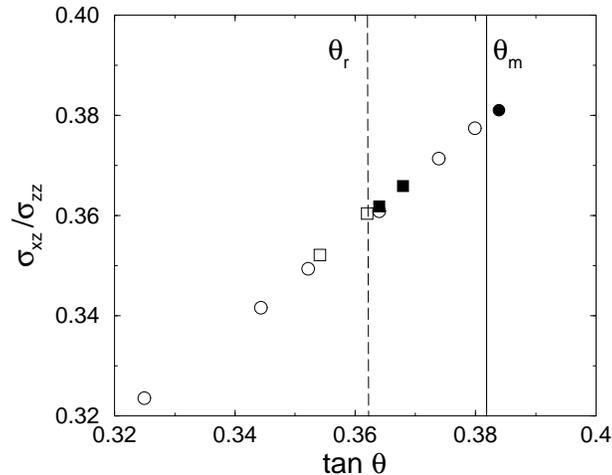}
\end{center} 
  \caption{Ratio of the shear stress to the normal stress
    $\sigma_{xz}/\sigma_{zz}$ as a function of the tangent of the
    angle of inclination $\theta$ for initiation of flow (squares) and
    cessation of flow (circles). The open symbols correspond to static
    states and the filled symbols are flowing. The lines indicate the
    positions of $\theta_{r,m}$. The stress ratio is linear with
    $\tan\theta$, but flow will not occur unless the stress ratio
    remains above $\tan\theta_{r}$ in the case of a system already
    flowing, or exceeds $\tan\theta_{m}$ for initiation of flow.}
  \label{chute3_fig16} 
\end{figure}

Another feature of flow transitions is in understanding the stress
response of the system upon initiation and cessation of flow.
Classical treatments of granular media \cite{nedderman1} demand that
failure will only occur if the ratio of the shear stress to the normal
stress equals the tangent of the angle of repose,
\begin{equation}
\sigma_{xz} = \sigma_{zz} \tan\theta_{r}.
\label{chute3_eq10}
\end{equation}
However, Eq.~\ref{chute3_eq10} needs to be modified to take into
account the history dependence of the starting point of flow for
initiation at $\theta_{m}$, and the general stopping point of flow for
cessation at $\theta_{r}$. In Fig.~\ref{chute3_fig16} the stress ratio
$\sigma_{xz}/\sigma_{zz}$ is plotted as a function of inclination
angle $\theta$ for initiation of flow (circles) and cessation of flow
(squares). We see that $\sigma_{xz}/\sigma_{zz}$ is linear with
$tan\theta$. Still, to observe flow from initiation, the stress ratio
must exceed $\tan\theta_{m}$, or for flow to cease its value must be
reduced to below $\tan\theta_{r}$. In other words, the shear stress
can have a non-zero value and remain static as long as the stress
ratio is smaller than the tangent of $\theta_{r}$ or $\theta_{m}$.
The particular stress state of the {\it static} system is very much
history dependent and comes about from the imposed conditions at the
instant the system became static \cite{leo10}.

\section{Conclusions} 
\label{chute3_conc}
We have performed a systematic simulation study of inclined plane,
gravity-driven flow of granular materials, elucidating the reasons
behind the discrepancies that exist between several different
experimental procedures. These arise as a consequence of how far the
system is from jamming, or how close the inclination angle of the pile
is from the angle of repose.

We have accurately determined the flow phase diagram for a system with
a particular interaction parameter set, indicating the height
dependence of the angle of repose and the maximum angle of stability.
We have also determined the boundary between steady state and unstable
flow and find that the maximum angle for steady state flow also
exhibits height dependence. This suggests the existence of an
intrinsic length scale in granular flows and forms part of our work in
progress.

For thick enough piles flowing on a rough inclined plane, say
$h\gtrsim20d$, the velocity profiles and rheology follow Bagnold
scaling. As one reduces the height of the flowing pile, a continuous
transition from Bagnold rheology to linear velocity profiles through
to avalanche-like dynamics occurs, until finally, one reaches the
angle of repose $\theta_{r}$, and flow ceases. As a result, one can
obtain various velocity profiles and flow behaviors not only through
changing the height of the flowing pile, but also by varying the
inclination angle of the chute, such that there is an overlap region
where one can obtain similar flow properties through either procedure.
These features can be used in surface flow treatments
\cite{degennes5,daerr3} to better predict the evolution of an
avalanche surface.

Our investigation on the initiation and cessation of flow shows that
the initiation process starts from failure at the base that results in
a ``heat'' front that propagates up the pile. Upon cessation of flow,
an effective freezing front propagates up the pile with particles
becoming trapped in the frozen layer until the whole system comes to
rest. In some cases, flow initiation or cessation only occurred after
many time units (approximately equivalent to 20 seconds for a 1mm
sized particle). A consequence of these findings is that experimental
reports of various flow behaviour in flowing piles may also be due to
the time window of observation. In general, we have captured all the
major observations from experiment, including quantitative agreement
of the scaling parameter $\beta$, Eq.~\ref{chute3_eq9}, from
Pouliquen's empirical theory \cite{pouliquen1}.

\section*{Acknowledgements}
We thank Deniz Ertas and Thomas Halsey for helpful discussions and a
critical reading of the manuscript. LES would also like to thank
M.~A.~Aguirre for insightful discussions. This work was supported by
the Division of Materials Science and Engineering, Office of Science,
U.S. Department of Energy. Sandia is a multiprogram laboratory
operated by Sandia Corporation, a Lockheed Martin Company, for the
United States Department of Energy under Contract DE-AC04-94AL85000.


\begin{thebibliography}{10}
\bibitem{footnote12}
High polydispersity may lead to such effects as segregation.
\end{thebibliography}
\end{document}